# Metasurface Freeform Nanophotonics

*Alan Zhan [1], Shane Colburn [2], Christopher M. Dodson [2], Arka Majumdar [1,2,+]*

[1]Department of Physics, University of Washington, Seattle, WA-98195, USA

[2]Department of Electrical Engineering, University of Washington, Seattle, WA-98195, USA

[+] Corresponding Author: arka@uw.edu

**Abstract**:

Freeform optics aims to expand the toolkit of optical elements by allowing for more complex phase geometries beyond rotational symmetry. Complex, asymmetric curvatures are employed to enhance the performance of optical components while minimizing their weight and size. Unfortunately, these asymmetric forms are often difficult to manufacture at the nanoscale with current technologies. Metasurfaces are planar sub-wavelength structures that can control the phase, amplitude, and polarization of incident light, and can thereby mimic complex geometric curvatures on a flat, wavelength-scale thick surface. We present a methodology for designing analogues of freeform optics using a low contrast dielectric metasurface platform for operation at visible wavelengths. We demonstrate a cubic phase plate with a point spread function exhibiting enhanced depth of field over 300 μm along the optical axis with potential for performing metasurface-based white light imaging, and an Alvarez lens with a tunable focal length range of over 2.5 mm with 100 μm of total mechanical displacement. The adaptation of freeform optics to a sub-wavelength metasurface platform allows for the ultimate miniaturization of optical components and offers a scalable route toward implementing near-arbitrary geometric curvatures in nanophotonics.

The function of an optical element is intrinsically tied to its geometry. While manufacturability has often constrained optical elements to have rotational invariance, the emerging field of freeform optics leverages more complex curvatures, often involving higher (> 2) order polynomials of the spatial dimension, to enable novel functionalities and simplified compound optical systems[1]. These elements have been shown to be capable of correcting aberrations[2], off-axis imaging[3], expanding field of view[4], and increasing depth of field[5]. Recent interest in freeform optics has been driven by potential applications in near-eye displays[6, 7] as well as compact optical systems for medical, aerospace, and mobile devices where there are stringent constraints on the size and weight of the optical package[8]. One surface of particular interest is the cubic profile, where the surface of the optical element is defined by a cubic function. These elements have been shown to exhibit increased depth of focus[9, 10], and in tandem, they can form an aberration-correcting lens with adjustable focus called the Alvarez lens[11, 12]. Many methods of realizing freeform optical elements, and in particular cubic surfaces, have been suggested and implemented, including fluid-filled[13], custom single-point diamond turned polymer[14], and diffractive optical elements[15]. Unfortunately, the thickness of these optical elements is can be variable and in general are larger, resulting in an increased overall volume. Unlike conventional optics, metasurface optical design is curvature agnostic, readily accepting both conventional spherical curvatures as well as complex freeform surfaces onto a flat form factor with no additional design difficulties. Moreover, well-developed semiconductor nanofabrication technology can be readily employed to fabricate such structures.

Metasurfaces are two-dimensional arrays of sub-wavelength scale scatterers arranged to arbitrarily control the wavefront of incident electromagnetic waves[16, 17]. Rather than relying upon gradual phase accumulation, metasurfaces impart an abrupt, spatially varying phase profile on the incident light. This allows us to map complex curvatures onto a flat, wavelength scale thick surface by converting them into a discretized spatial phase profile. In addition to their compact size and weight, metasurfaces are fabricated using a single step lithography procedure with mature, highly scalable nanofabrication technology developed by the semiconductor industry. Numerous different metasurface material platforms have been demonstrated, including noble metals[17-19], high contrast dielectrics[20, 21], and low contrast dielectrics[22, 23]. For visible wavelengths, low contrast dielectrics, such as silicon nitride, are desirable as they do not suffer from absorption losses due to their wide band gap and also exhibit similar performance to other material platforms. In recent years, all dielectric metasurfaces have been used to build many different optical components such as quadratic lenses, vortex beam generators, and holograms[17-20]. However, there has been little research in realizing freeform optical elements in visible frequency for imaging applications utilizing a metasurface platform. While both vortex beam generators and holograms lack rotational symmetry, their spatial phase functions are not characterized by higher order polynomials, ( > 3) as is the case for most freeform optics. In this paper, we present a silicon nitride metasurface-based cubic phase optical element and an Alvarez lens operating at visible wavelengths. We observed an extended depth of focus (~300 μm), enough to ensure an identical point spread function (PSF) for red and green light at the same image plane, potentially enabling white light imaging. Additionally, we experimentally demonstrated a change in focal length of ~2.5 mm by a physical displacement of only 100 μm using the Alvarez lens. This is the highest reported focal length tuning range in metasurface optics, and most importantly, the change in the focal length is significantly larger (~25 times) than the actual physical displacement.

In our metasurface design process, we take the sag profile of an arbitrary freeform surface, described by its height *(z)* as a function of its in-plane coordinates *(x, y)* as in Fig 1a, and convert it into a discrete phase profile. We then quantize the phase profile into six linear steps from 0 to 2π corresponding to cylindrical posts with diameters *d* ranging from 192 nm to 420 nm using the corresponding values shown in Fig 1b. We choose a set of parameters for posts with thickness $t = \lambda$, in this case 633 nm, arranged on a square lattice with periodicity $p = 0.7\,\lambda$, or 443 nm, (Figs 1c, d). Due to the discretization of the phase profile, there is a fundamental limitation on achievable curvatures for any specific sampling periodicity based on the Nyquist-Shannon sampling theorem:

$$\Lambda_s < \frac{\pi}{|\nabla \varphi(x,y)|_{max}}, \quad (1)$$

where $\Lambda_s$ is the sampling periodicity, and $\varphi(x,y)$ is the spatial profile to be sampled. This criterion ensures an accurate sampling of an arbitrary spatial phase profile. A derivation of this limitation, and its effect on device parameters is provided in supplementary material S5.

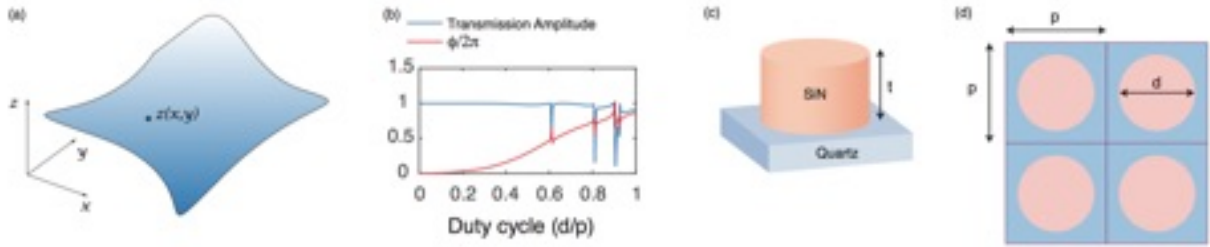

Figure 1: Mapping a freeform surface onto a metasurface: An arbitrary freeform surface is shown in (a). The corresponding height z(x,y) is converted into a discretized phase profile using the pillar parameters shown in (b). The parameters in (b) are capable of producing a full cycle of phase shifts and also maintain large regions of continuous, near unity transmission amplitude. (c) and (d) are simple schematics of a metasurface with thickness t, periodicity p, and diameter d.

Cubic phase elements have been explored for wave-front coding as part of a focus-invariant imaging system[9, 10]. These cubic phase elements do not cause incident light to converge into single point; instead incident rays converge along an extended length of the optical axis, allowing the point spread function (PSF) of the element to remain relatively constant for a large range of displacements along the optical axis. The images produced by such systems are often unintelligible to the human eye, but they can be digitally post-processed using knowledge of the cubic element's PSF to recreate an image with enhanced depth of focus. More detail on the deconvolution process for the image is provided in the supplement S8. We design a cubic element with the phase profile:

$$\varphi(x,y) = mod\left(\frac{\alpha}{L^3}(x^3 + y^3), 2\pi\right), (2)$$

where (x,y) are the device's in plane coordinates, $L$ is the width of the design, and $\alpha$ is a constant determining the rate of the phase variation on the metasurface. Larger values of $\alpha$ lead to better depth invariance at the expense of increased noise in the image while small values compromise the depth invariance[24]. Motivated by previous designs, we choose a value of $\alpha = 14\pi$[24]. For our choice of parameters, the sampling periodicity $p$ is an order of magnitude smaller than that of the limit, satisfying the criterion in (1).

The Alvarez lens is a compound optical element consisting of two cubic phase plates with one obeying the phase profile:

$$\varphi_{alv}(x,y) = mod(\frac{2\pi}{\lambda} A \left(\frac{1}{3}x^3 + xy^2\right), 2\pi), (3)$$

and the other obeying its inverse such that $\varphi_{alv}(x,y) + \varphi_{inv}(x,y) = 0$, where (x,y) are the device's in plane coordinates, and $A$ is a constant determining the rate of phase variation on the metasurface. If the two elements are perfectly aligned, the Alvarez lens does not focus light, which can be interpreted as there being a focal length of infinity. Laterally displacing the elements relative to each other along the x-axis allows us to focus at finite lengths. Moreover, by controlling the extent of the lateral displacement along the x-axis we can change the focal length. Larger values of $A$ increase the range of tunable focal lengths at the expense of image quality[12]. The range focal length with respect to displacements is given by the expression[11, 12]:

$$f = \frac{1}{4Ad}, (4)$$

where $f$ is the focal length, $A$ is the same constant as in the phase profile and $2d$ is the relative displacement of the two surfaces meaning the Alvarez lens is displaced by a distance $d$ and the inverse lens is displaced by $-d$ from the origin. A derivation of the focal length expression is provided in the supplementary material S2. We emphasize that, unlike changing periodicity by stretching a metasurface lens[25], this method can provide a much larger change in the focal length. Our choice of parameters for the Alvarez lens is also within the limit of the criterion (1).

We fabricated a cubic metasurface with $\alpha = 14\pi$ and L = 150 μm, and a set of square Alvarez metasurfaces with $A = 1.17 \times 10^7$ m$^{-2}$, and length 150 μm. The devices are fabricated in 633 nm silicon nitride deposited on top of a 500 μm fused quartz substrate. Scanning electron micrographs (SEMs) of the finished devices coated in gold are shown in Fig 2.

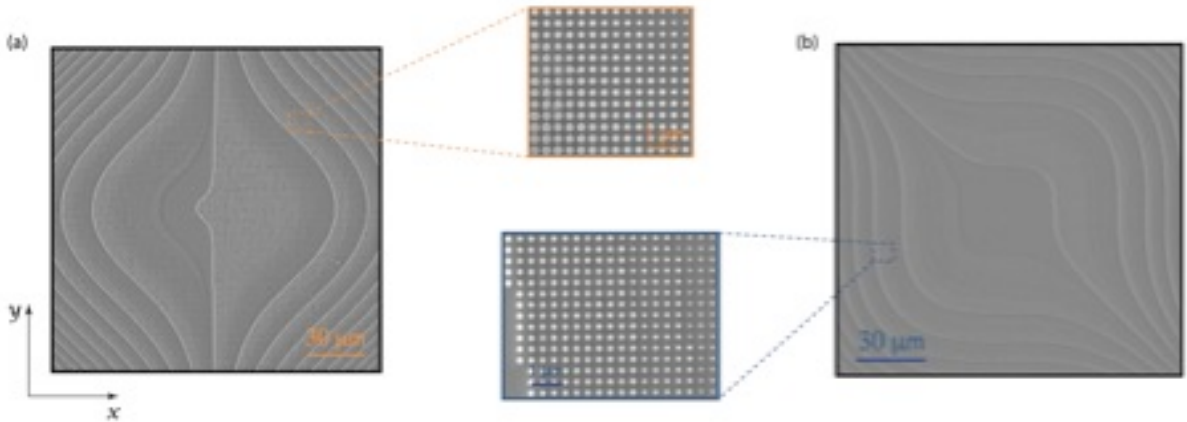

Figure 2: Scanning electron micrographs of fabricated devices coated in gold. Half of the Alvarez lens is shown in (a), and the cubic phase plate is shown in (b). Insets are zooms of specific locations of the metasurface showing the gradient in pillar sizes.

The cubic metasurface is characterized using a microscope free to translate along the optical axis. Note that the phase plates are designed to function with incoherent illumination[9, 10], but the power of our LEDs was not high enough to determine the PSF. The cubic phase plate measurements were performed on a setup shown in supplementary Fig S2. Light from a helium-neon laser was coupled to a fiber for the red measurements, and light from a 532 nm laser was used for green measurements. The light was sent through a 5 μm fixed pinhole (Thorlabs P5S) before illuminating the sample mounted on a standard 1 mm glass microscope slide with the metasurface facing the microscope. The cubic PSFs were measured using 4 mW of power incident on the pinhole, and the lenses' PSFs were measured using 1.5 mW of incident power. A home-built microscope comprising of a 40x objective (Nikon Plan Fluor) with a working distance of 0.66 mm and NA 0.75 and a tube lens (Thorlabs ITL200) with a focal length of 20 cm is used to measure the field profiles. This microscope images the intensity profile generated by the cubic phase plate onto a Point Grey Chameleon CCD. The magnification of the setup was determined using known dimensions of the fabricated metasurface. By translating the microscope along the optical axis (z) we were able to image the intensity profile in steps of 25.4 μm to capture the images shown in Fig 3 with respect to the z displacement. We see that indeed, the PSF of the element changes minimally with displacements along the optical axis of over 300 μm, confirming the depth-invariant behavior of the cubic phase plate. The slight discrepancy in the PSF is primarily due to experimental noise. Hence, in addition to the measurement of the PSF, we also calculated and compared the modulation transfer function (MTF), shown in supplement S8. The calculated MTFs are very similar and for comparison, we also measured the PSF of a metasurface lens (quadratic phase profile) with a focal length of 500 μm shown in Fig 3c, d. It is clear that the PSF of the lens is highly dependent upon displacements along the optical axis, changing substantially over a range of 100 μm, unlike that of the cubic phase plate. While the cubic metasurface exhibited a large range of displacements for which the PSFs were similar for the two illumination wavelengths (red and green), the metasurface lenses exhibit significant chromatic aberrations. With the understanding that an image is the convolution of an object with the imaging system's impulse response or PSF, this effect could be exploited for performing white light imaging. If the PSF is identical for a range of wavelengths, deconvolution of the image can be performed with a single digital filter obtained from the imaging system's PSF[26]. For highly chromatic optical elements, this is not possible as shown in Fig 3c, d, but we can utilize the cubic element's increased depth of focus to find a point where the PSF is the same for a range of wavelengths. This may truly enable broadband operation, unlike previously reported results, where the lens only works for certain discrete wavelengths[27, 28].

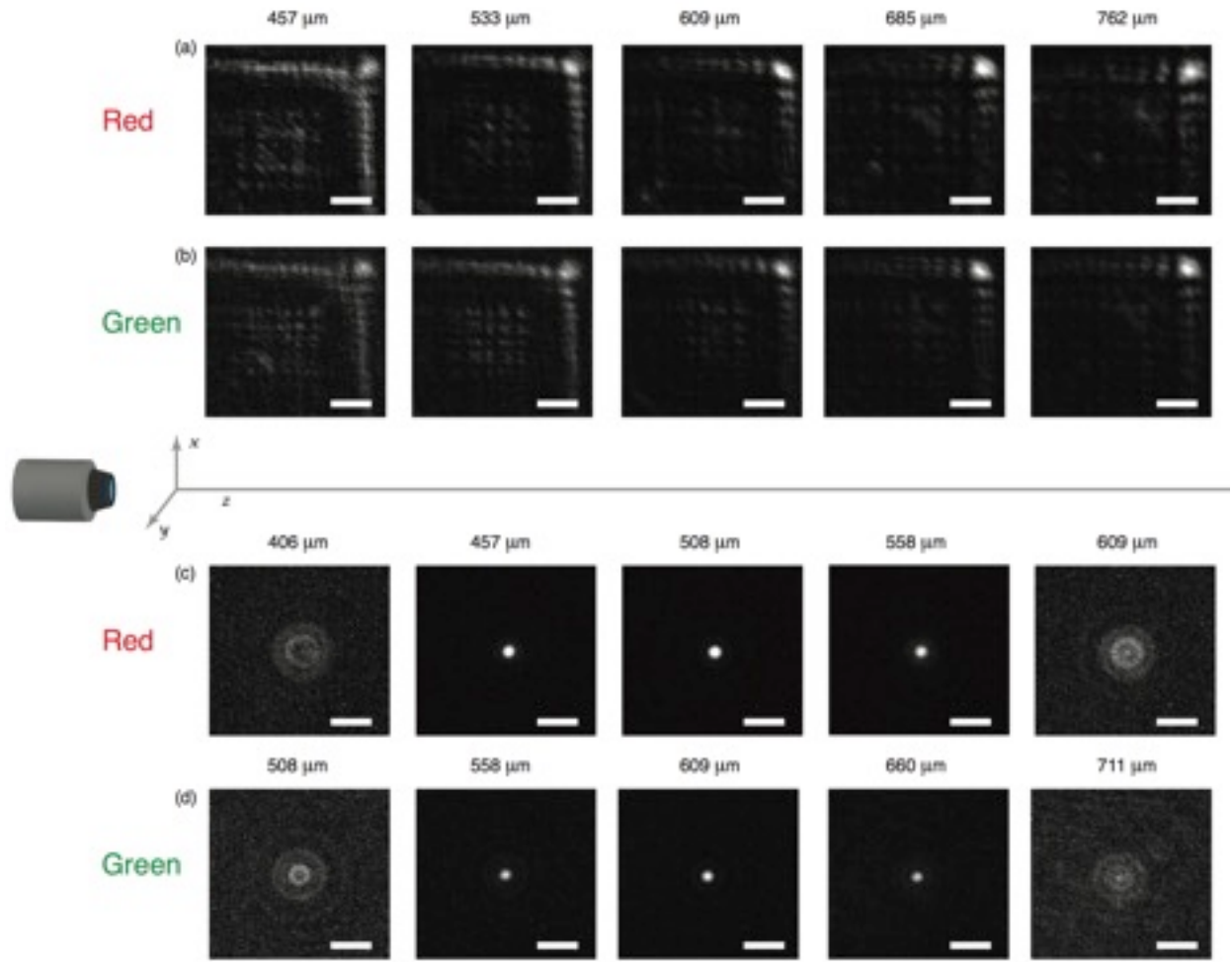

Figure 3: Dependence of cubic metasurface and metasurface lens point spread functions (PSF) upon displacement along the optical axis. (a) and (b) are the PSFs of the cubic element under coherent illumination by red and green light respectively. (c) and (d) are the PSFs of a 500 μm metasurface lens from ref [22] under red and green illumination. All figures share the same 18μm scale bar. While the intensity profiles at 558 μm seem similar for the quadratic lens (c) and (d) seem similar, analysis of their MTF shown in supplement S9 shows a significant difference.

The performance of the Alvarez lens was measured using a setup shown in supplementary Fig S3. Red light is obtained from a fiber-coupled light-emitting diode (Thorlabs M625F1) and directed towards the sample. The Alvarez lens consists of the Alvarez phase plate and the inverse phase plate, and the two samples are mounted with the devices facing each other. The Alvarez phase plate is mounted on a standard 1 mm glass slide while the inverse phase plate is mounted on a thin glass coverslip with a thickness between 0.16 to 0.19 mm (Fisherbrand 12-544-E). The Alvarez phase plate is placed on the illumination side while the inverse phase plate is placed on the microscope side. Finally, the Alvarez phase plates were mounted on an x-z translation stage enabling control over the displacement between the two phase plates in the x and z directions. The x direction can move in increments as fine as 0.5 μm. The focal distance of the Alvarez lens is measured for displacements of 2 μm to 50 μm in steps of 2 μm. For each displacement, the microscope is translated along the z axis, imaging intensity profiles in steps of 25.4 μm. Due to the sensitivity of the focal length to small misalignments, all data was taken consecutively from one displacement to the next with one alignment at the beginning of the measurement. Measurements for five displacement values showing the microscope moving into and out of the focal plane are shown in Fig 5c, d. The alignment of the two metasurfaces is done one at a time by first imaging the first metasurface on the CCD and marking a single corner with a marker. The microscope is then translated backwards along the optical axis to allow us to bring the second metasurface into focus and translate its corner to the same marker. Finally, the two metasurfaces are translated along the optical axis to minimize their separation by eye. In order to minimize the separation between the two elements, both Alvarez lenses were mounted on stages free to move along the optical axis. The distance between the two was determined using the microscope by focusing on each element separately and recording their positions. The two elements were then brought together to their final separation of less than 0.3 mm. We did not bring the elements closer because of the possibility of scratching the elements. Simulation and experimental data on the axial separation between the plates is presented in supplement S7. Illustrations of the behavior of the Alvarez phase profile for displacements along the x axis are shown in Fig 4. For small displacements, the resulting phase profile is slowly spatially varying, corresponding to a lens with a large focal length, while large displacements correspond to a highly varying phase profile, or a short focal length lens. The theoretical performance of the lens based on the previous formula for our design parameters is shown in Fig 4j.

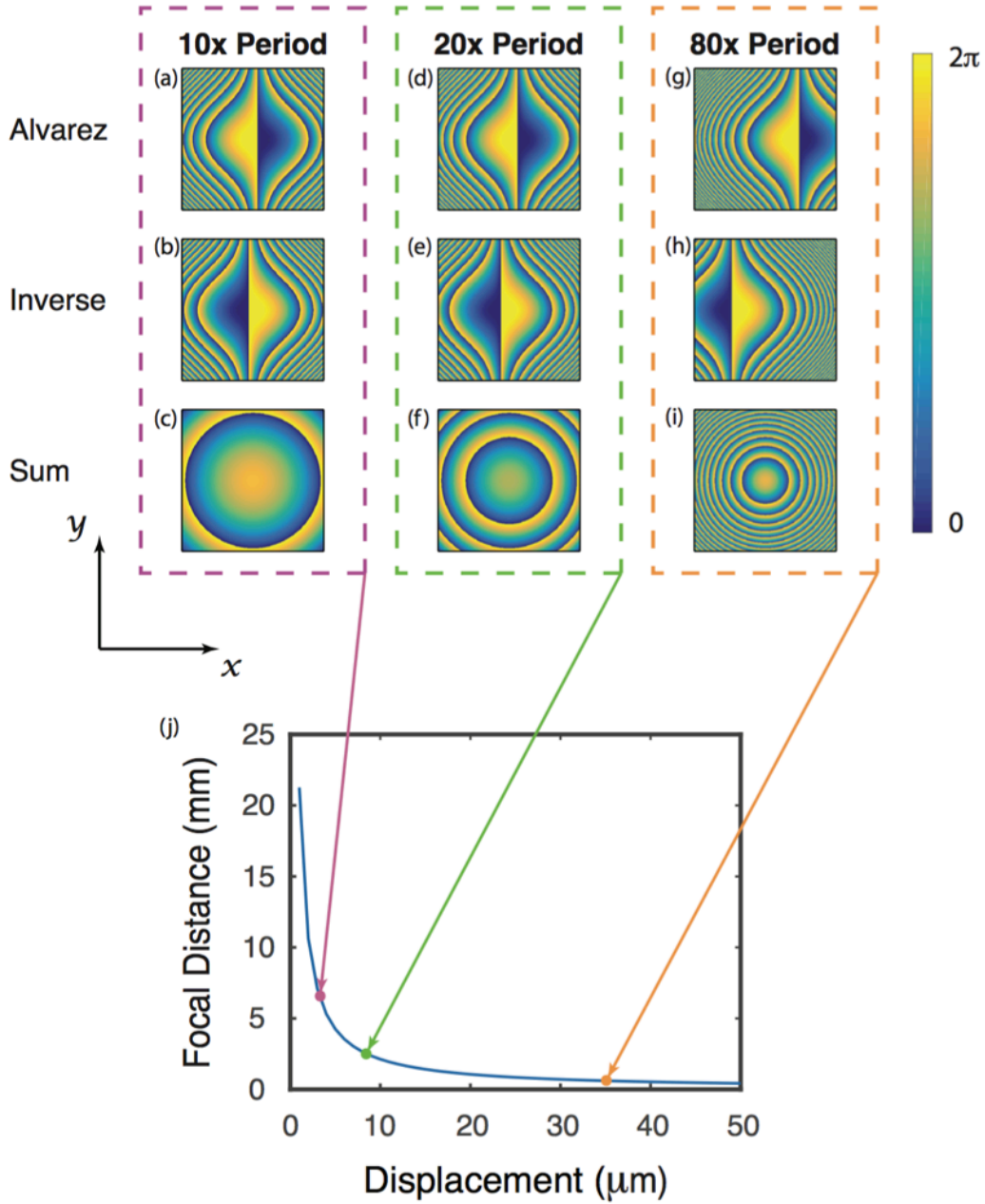

Figure 4: Behavior of the Alvarez lens in response to x displacement. (a), (d), (g) represent the phase profiles of one Alvarez element for displacements of 10p, 20p, 80p respectively, (b), (e), (h) represent their inverses at displacements of −10p, -20p -80p respectively, and (c), (f), (i) are the sums of the displaced phase profiles. The phase profiles are displaced in units of the metasurface lattice periodicity p = 443 nm, with (a)-(c) representing a 4.43 μm displacement, (d)-(f) representing 8.86 μm displacement, and (g) – (i) representing a 35.4 μm displacement. (j) Plot of focal length dependence on displacement based on equation 4. Larger displacements result in a more rapidly varying phase profile, corresponding to a lens with a smaller focal length. The colored dots indicate the focal lengths of lenses shown in (c), (f), (i). Parameters used are the same as for the fabricated device, L = 150 μm, A = $1.17 \times 10^7$ m$^{-2}$.

We have experimentally measured the focal lengths for displacements *d* of each metasurface from 2 to 50 μm and find the focal distances change from a minimum of 0.5 mm to a maximum of 3 mm as seen in Fig 5a. This indicates that with a physical displacement of 100 μm, the focal length changes by 2.5 mm. While the experimentally measured change in the focal length is significantly smaller than the simple theoretical predictions in Fig 4i, this change is still the largest among all the demonstrated changes in focal length by mechanically actuated metasurface-based tunable optical elements[25, 29, 30]. In addition, we emphasize that the lens achieves most of its focal tuning range at a small range of physical displacement, in that we can tune the focal length by 2 mm using only around 30 μm of physical displacement. We performed a simple fit of the form:

$$f(\text{d}) = \frac{1}{4A(\text{d}+B)}, \quad (5)$$

to generate the red line shown in Fig 5a. The best fitting parameters are A = $7.97 \times 10^6$ m$^{-2}$, similar to our design value of $1.17 \times 10^7$ m$^{-2}$ and B = 7.6 μm, which indicates the extent of misalignment. We believe the major sources of the discrepancy between the measurement and the theoretical prediction are this small degree of misalignment (of order B) and also the discretized phase profile of the metasurface, in contrast to the continuous profile assumed in the theory. The effect of discretized phase is verified via FDTD simulations of a metasurface-based Alvarez lens presented in the supplementary material S1. Previous focus-tunable metasurface lenses were based on stretchable substrates, which have a focal length dependence $f \propto (1 + \epsilon)^2$ in the paraxial limit where $\epsilon$ is the stretching factor[25], corresponding to a change $\Delta f = (2\epsilon + \epsilon^2)f$. This change is linear to first order in $\epsilon$ with the quadratic term dominating for greater than unity stretch factors, whereas the change in focal length of the Alvarez lens behaves nonlinearly as shown in the equation (3) (details provided in the supplementary material S2), with the largest changes in focal length occurring for the smallest physical displacements. Another important quantity to assess the quality of a lens is the spot size, which we measure by calculating the full width at half maximum (FWHM) of a Gaussian fit to a 1D slice of the intensity data. The FWHM shows a similar dependence on lateral displacement as the focal length. The largest focal length of ~3 mm displays the largest FWHM of ~20 μm, while the smallest focal length of ~0.5mm has a FWHM of ~5 μm (Fig 5a). We find that our measured FWHM is near diffraction-limited using the methodology in ref. 22 (Fig 5b). In addition, we characterize the behavior of the lens as it moves into and out of the focal plane as shown in Fig 5c, d. The FWHM of the lens is measured using a horizontal (x) and vertical (y) 1D cross-sections for Fig 5c, d, respectively. Mirroring the results from our numerical simulations (supplementary materials S1, S4), the beam spot is wider along the x than along the y axis.

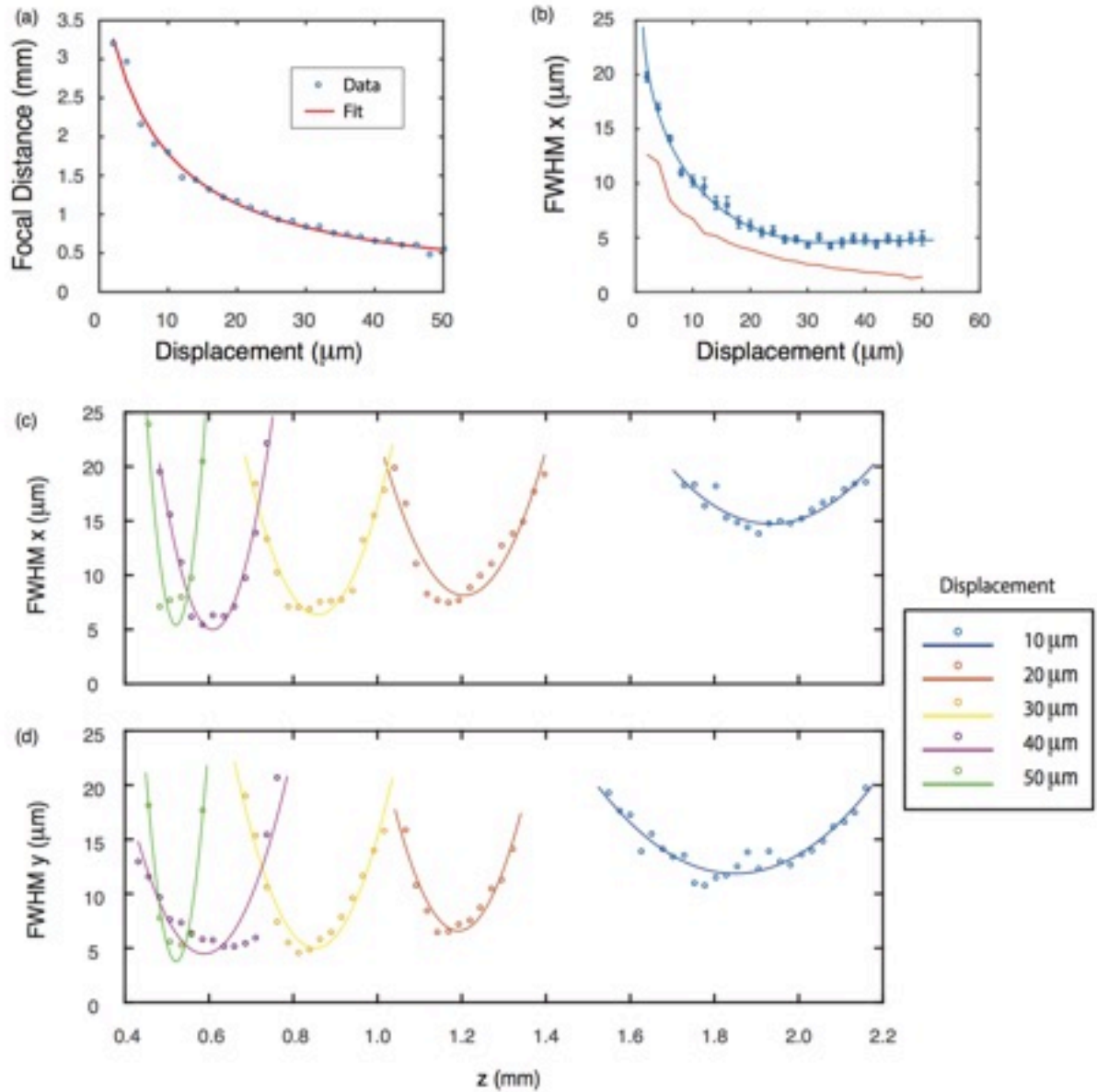

Figure 5: Alvarez lens performance. (a) Measured focal distance of the Alvarez lens pair plotted against x displacement. The red line is a theoretical fit to the focal length data. (b) Full width at half maximum (FWHM) measured along the x axis plotted against x displacement. The measured data are shown as blue points while the red line is an eye guide. The diffraction-limited spot size FWHM is plotted in red. Error bars represent a 95% confidence interval of a Gaussian fit. For both (a) and (b) images were taken with a displacement step size of 2 μm. (c), (d) Behavior of the Alvarez lens FWHM for five displacements along x-axis. The FWHM of the spot-size in the sensor plane is plotted as the microscope moves into and out of the focal plane. The FWHMs are measured along the (c) x and (d) y axes. FWHM data is plotted as the points, and the lines are eye guides.

We have fabricated and demonstrated the performance of a metasurface-based cubic phase element and Alvarez lens in silicon nitride. To the best of our knowledge, these are the first metasurface-based optical elements designed using the principles of freeform optics. We believe this metasurface platform is near ideal for both adapting existing freeform optical elements, and also realizing new classes of arbitrary spatial phase profiles provided the Nyquist-Shannon sampling criterion is satisfied. This platform also has the unprecedented ability for the integration of freeform optical elements at the micron scale leading to ultra-miniature optical systems. For example, throughout tunable optical designs, we find that a mechanical change of $x$ nm results in a change in focal length or in resonance wavelength of the order $O(x)$ nm[31, 32]. In the case of the Alvarez lens no such limitations exist, and we demonstrated greater than 2 mm focal length tuning, with only tens of microns of physical displacement. Such a small displacement is beneficial, especially if the displacement is realized using integrated MEMS devices. Similarly, by using a non-quadratic phase profile, we can realize white light imaging in diffractive optics. In particular, our results indicate a depth-invariant point spread function for red and green lasers for the cubic phase-mask, resulting in the same PSF for both colors at the image plane. The reported metasurfaces involving cubic phase profiles represent a first step towards the promising new field of metasurface-enabled freeform optics, which will find applications in correcting aberrations, building compact optical systems or sensors, such as realizing near-eye displays[6, 7] or ultra-compact endoscopes[8]. Additionally, by adapting existing semiconductor technologies, such as nano-imprint lithography, these devices can easily be fabricated in a scalable manner.

ASSOCIATED CONTENT

**Supporting Information**. Simulation results for the performance of the Alvarez lens, a derivation of the focal length of the Alvarez lens, the experimental setups, a criterion for diffraction limited spots, a calculation of the achievable curvatures based on the Nyquist-Shannon theorem, the chromatic behavior, the dependence of performance upon axial separation, and a calculation of the modulation transfer function. This content is available free of charge through the internet at http://pubs.acs.org.

AUTHOR INFORMATION

**Author Contributions**

C.M.D. and A.M. conceived the idea. A.Z. and S.C. performed the numerical simulations and design. A.Z. fabricated the device, performed the experiment, and wrote the paper with input from everyone. A.M. supervised the whole project.

**Funding Sources**

The research work is supported by the startup fund provided by University of Washington, Seattle, and an Intel Early Career Faculty Award.

**Acknowledgements**: All of the fabrication was performed at the Washington Nanofabrication Facility (WNF), a National Nanotechnology Infrastructure Network (NNIN) site at the University of Washington, which is supported in part by the National Science Foundation (awards 0335765 and 1337840), the Washington Research Foundation, the M. J. Murdock Charitable Trust, GCE Market, Class One Technologies, and Google. The research work is supported by the startup fund provided by University of Washington, Seattle, and an Intel Early Career Faculty Award. We acknowledge Dr. Andrei Faraon for helpful discussion and constructive feedback during the preparation of the manuscript.

**Table of Contents Figure:**

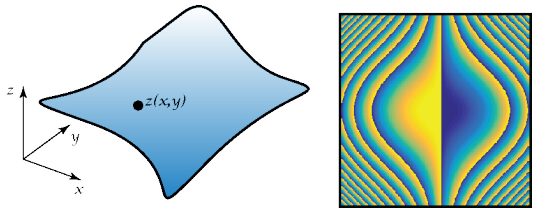

# Metasurface Freeform Nanophotonics

*Alan Zhan [1], Shane Colburn [2], Christopher M. Dodson [2], Arka Majumdar [1,2]*

[1]*Department of Physics, University of Washington, Seattle, WA-98195, USA*

[2]*Department of Electrical Engineering, University of Washington, Seattle, WA-98195, USA*

15 pages, 9 figures S1-S8

**Supplement: Metasurface Freeform Nanophotonics**

**S1: Simulation Results**

We performed finite-difference time-domain (FDTD) simulations of the metasurface-based Alvarez lens to understand the effect of discretization of the phase profile. We find the change in the focal length qualitatively matches the theoretical predictions assuming a continuous phase profile, but the numerically calculated focal lengths do not quantitatively match well with the theoretical equation derived for a continuous phase profile. In particular, the focal lengths deviate significantly at small displacement, as we also observed in our experiment. Additionally, we find the focal spot size is larger in the x direction than in the y direction, also in accordance with experiment. In calculating the diffraction limit for the x direction, we account for an increase in the physical lens size due to the displacement along that axis. This accounts for the differences in diffraction limits shown in Fig S1c for the x and y directions.

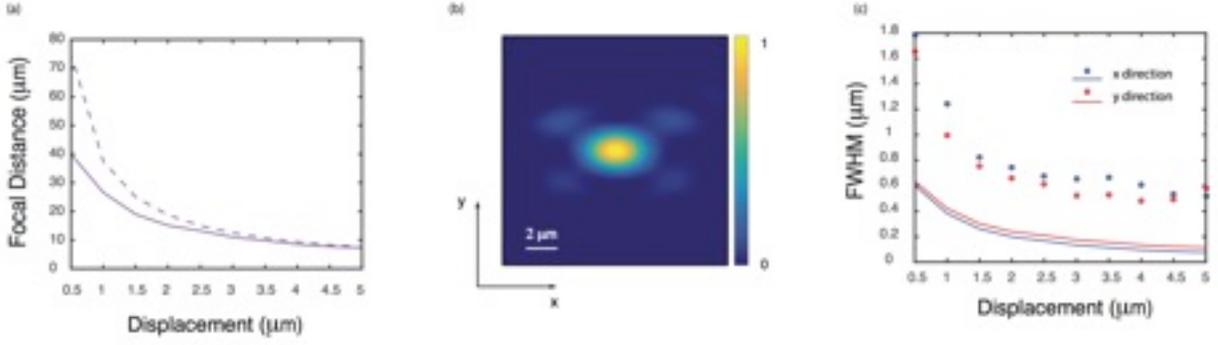

Fig. S1: FDTD simulation results for an Alvarez lens. (a) the measured focal length plotted against lateral displacement. The simulation data is shown as the solid line, and the theoretical focal length range (assuming a continuous phase profile) is shown as the dotted line. Displacements are made in steps of 0.5 μm. (b) an example of a simulated focal spot for a 0.5 μm displacement. (c) The numerically estimated FWHM for each displacement step of 0.5 μm. The x and y FWHM are plotted as points that are blue and red respectively. The calculated diffraction limit corresponding to the x and y geometric parameters of that lens are shown as solid lines in blue and red respectively. Parameters for simulation are A = 6.67×10$^9$ m$^{-2}$, and the phase plates are 10 μm x 10 μm.

## S2: Alvarez Focal Length Formula Derivation:

The central concept of the Alvarez lens is the dependence of the focal length on the lateral displacement of the two phase plates, the Alvarez phase plate obeying:

$$\varphi_{Alv}(x, y) = A\left(\frac{1}{3}x^3 + xy^2\right), (1)$$

and the inverse phase plate obeying its negative:

$$\varphi_{Inv}(x, y) = -A\left(\frac{1}{3}x^3 + xy^2\right), (2)$$

such that $\varphi_{Inv}(x, y) + \varphi_{Alv}(x, y) = 0$ for aligned phase plates. For a displacement $d$ along the x axis, the addition of the two surfaces produces a quadratic phase profile plus a constant phase offset:

$$\varphi_{Sum}(d) = \varphi_{Alv}(x+d,y) + \varphi_{Inv}(x-d,y) = 2Ad(x^2+y^2) + \tfrac{2}{3}d^3, \quad (3)$$

neglecting the constant phase offset, and setting $r^2 = (x^2 + y^2)$, we recognize the expression for a lens under the paraxial approximation:

$$\varphi_{Lens}(d) = 2Adr^2 = \frac{r^2}{2f}, \quad (4)$$

with focal length as a function of displacement:

$$f(d) = \frac{1}{4Ad}, \quad (5)$$

**S3: Setups**

The experimental setups are shown in Fig. S2 and S3.

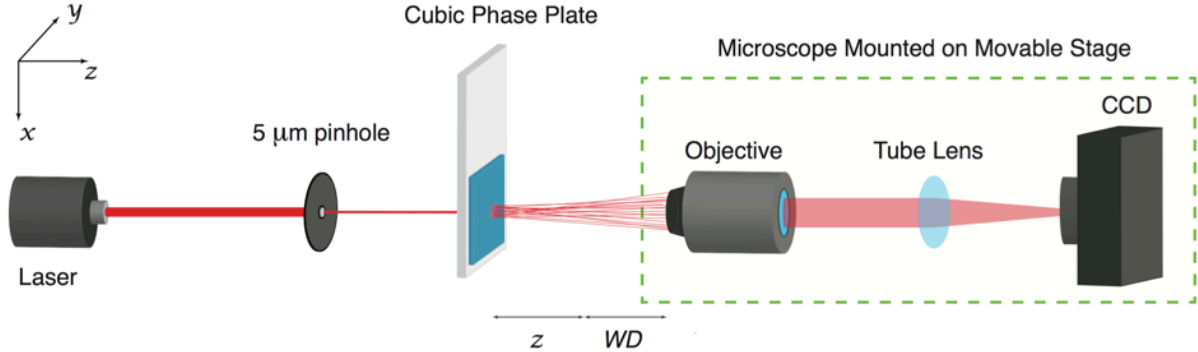

Fig. S2: Point spread function measurement setup: Schematic of the setup used to measure the point spread functions of the cubic metasurface phase plate and the metasurface lens. Illumination is provided either by a helium-neon laser for red or a 532 nm laser for green, and is passed through a 5 μm pinhole to approximate a point source. The microscope is free to move along the z axis.

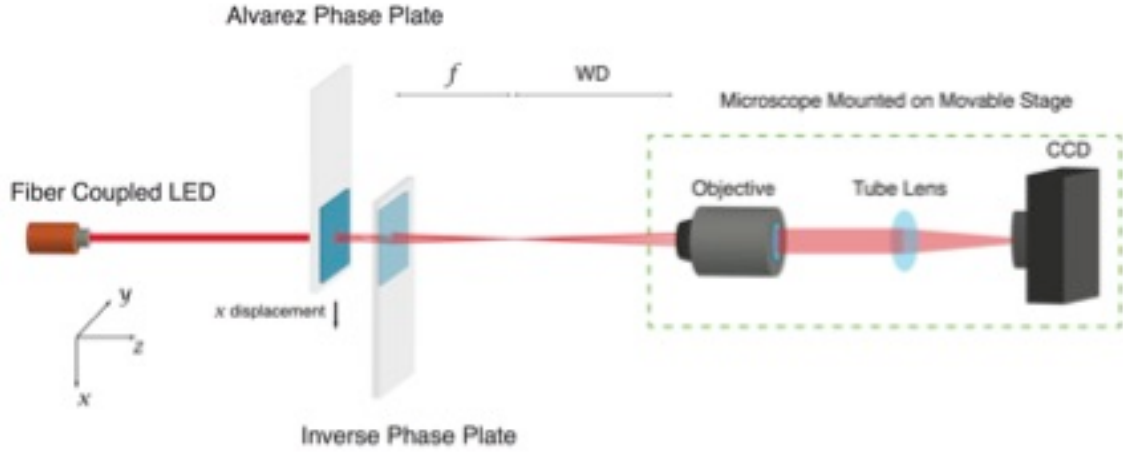

Fig. S3: Alvarez phase plate measurement setup: Schematic of the setup used to measure the performance of the Alvarez lens. Light is provided by a fiber-coupled red light-emitting diode (LED). The Alvarez phase plate is mounted on the LED side while the inverse phase plate is mounted on the microscope side. The Alvarez phase plate is allowed to move in the x direction. The microscope is free to move along the z axis, allowing us to image into and out of the focal plane for each displacement.

**S4: Measurement and Diffraction Limit**

The experimentally measured focal spot from the Alvarez lens shows different FWHM along x and y direction, which is consistent with the numerical FDTD simulations (Fig S1). Here we present our criterion for characterizing the focusing performances of a lens based on its FWHM. An ideal lens with focal length $f$ and radius $d$ will have an Airy disk intensity profile given by:

$$I(\theta) = I_o \left(\frac{2J_1(kd \sin \theta)}{kd \sin \theta}\right)^2$$

where $I_o$ is the central peak intensity, $J_1(x)$ is the first order Bessel function of the first kind, $k$ is the free space wave vector of the illuminating light, d is the lens radius, and $\theta$ is the angular position. The diffraction-limited FWHM for a particular lens with geometric parameters *f* and *d* is obtained by a Gaussian fit to the Airy disk intensity profile.

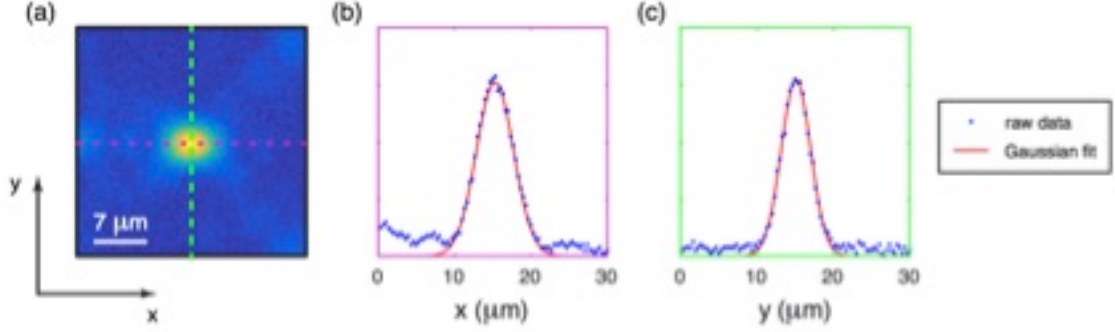

Fig. S4: Characterizing full width half maximum: (a) an example of an experimental focal spot for an Alvarez lens with 30 μm lateral displacement. A Gaussian fit is used along the x (b) and y (c) axes to estimate the focal spot size.

**S5: Fundamental Limitations**

In designing metasurface-based freeform optics, there are limitations on the kinds of phase functions that can actually be implemented. One of the fundamental limitations concerns how the continuously-defined phase function is spatially sampled by the metasurface's subwavelength lattice points. The Nyquist-Shannon sampling theorem requires that the function be bandlimited and that the sampling frequency $f_s$ be related to the maximum frequency component, as below, in order to prevent aliasing:

$$f_s = \frac{1}{\Lambda_s} > 2 f_{max}, \quad (6)$$

The highest frequency component can be related to the instantaneous frequency as below:

$$f_{max} = \frac{|\nabla \varphi(x,y)|_{max}}{2\pi}, \quad (7)$$

Solving for the sampling period gives:

$$\Lambda_s < \frac{\pi}{|\nabla \varphi(x,y)|_{max}}, (8)$$

Using (8) with a given metasurface periodicity which sets $\Lambda_s$, arbitrary phase functions can be tested and it can be determined whether or not it is possible to implement them. For example, in the case of freeform optics with phase functions consisting of higher order polynomials, there will be restrictions on the extent and functional form of $\varphi(x,y)$. Here we analyze two specific cases of $\varphi(x,y)$ to show the limitations of the metasurface optics.

For a parabolic lens with a phase profile given by (9), the spatial extent of the lens is limited to a maximum radius of $r_{max}$, and using (8) the restriction on $\Lambda_s$ is given by (10):

$$\varphi(r) = \frac{\pi r^2}{\lambda f}, (9)$$

$$\Lambda_s < \frac{\lambda f}{2 r_{max}}, (10)$$

Using (10) and setting $D = 2r_{max}$, the $D/f$ ratio can be determined to find the limitation on the NA given in (11):

$$NA < \sin\left[\arctan\left(\frac{\lambda}{2\Lambda_s}\right)\right], (11)$$

For a cubic profile given by (12), the restriction on $\Lambda_s$ is given in (13) where it is assumed the maximum value for both x and y is $L/2$:

$$\varphi(x,y) = \frac{\alpha}{L^3}(x^3 + y^3), (12)$$

$$\Lambda_s < \frac{2\sqrt{2}\pi L}{3\alpha}, (13)$$

This shows that for a given periodicity, we cannot have arbitrarily large $\alpha$, which dictates the depth of focus. This methodology is applicable to any arbitrary phase functions and can provide a baseline check for implementation feasibility for metasurface-based freeform optics.

**S6: Alvarez Chromatic Behavior**

The same lens as in supplement S1 was simulated at a displacement of 4 μm for wavelengths between 400 and 700 nm in steps of 50 nm. The electric field intensities in the x-z and y-z planes centered on the optical axis are plotted in Fig S6 for the range of simulated wavelengths. We find that the Alvarez lens fails to focus adequately at wavelengths below 550 nm, and displays expected chromatic aberrations in the wavelength range of 550 nm to 700 nm. At 400 and 450 nm, the wavelength is less than and approaching the periodicity respectively, so we do not expect the Alvarez lens to perform well in that regime. For wavelengths larger than the design, we expect the lens to perform adequately, although with the usual chromatic dispersion of diffractive optics.

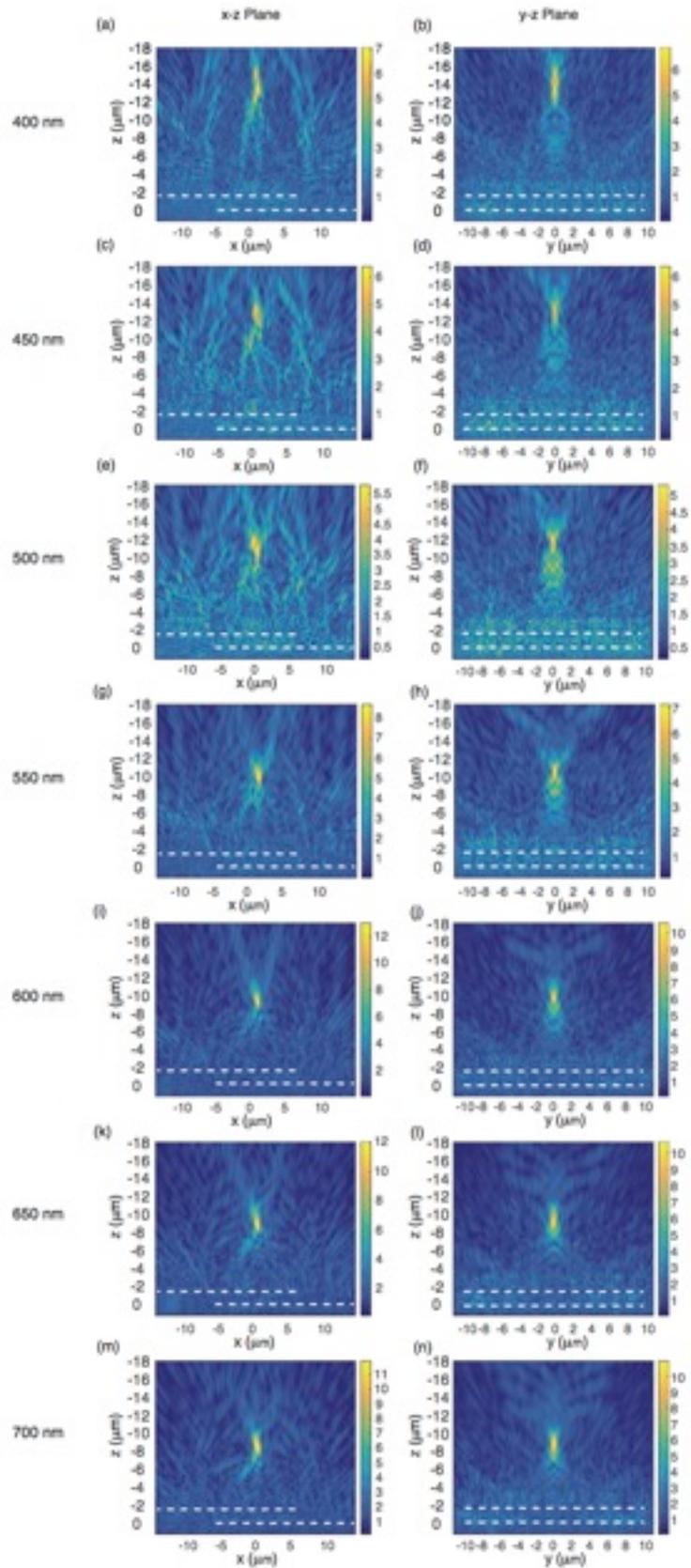

Fig S5: Chromatic behavior of the Alvarez lens. The electric field intensity profiles in the x-z and y-z planes are plotted, centered at the optical axis for illumination wavelengths covering the visible spectrum in steps of 50 nm. The lens begins to form a distinct focal spot for 550 nm in both the x-z and y-z planes. The white dashed lines indicate the locations of the two metasurfaces comprising the Alvarez lens.

**S7: Alvarez Axial Separation Behavior**

We investigated the dependence of the focusing behavior of the Alvarez lens on axial separation between the two metasurfaces via both FDTD simulations and experiment. Two Alvarez plates can be understood as generating Airy beams accelerating along a parabolic path on the axis of displacement (x for our design). As the axial separation between the metasurfaces increases, the initial Airy beam generated by the first metasurface begins to diverge from the second plate, which has finite extent, causing degradation of the focal spot.

In simulation, as shown in Fig. S6, axial separations have a large effect on the shape of the focal spot in the x-z plane, and also a large effect on the intensities of the focal spots for both planes. As seen in Fig. S6 (d), for large separations, the Airy beam generated by the first metasurface begins to clip the edge of the second.

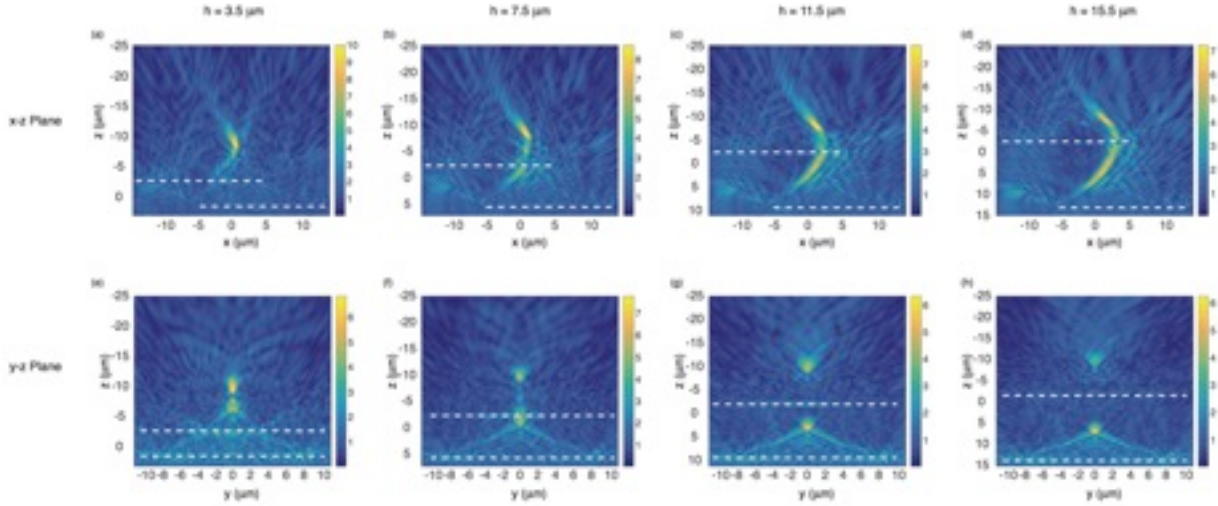

Fig. S6: Simulated Alvarez lens performance for separations along the optical axis. The electric field intensities are plotted. As the separation increases, the x-z plane focal spot deforms, elongating, and also decreasing in intensity (a)-(d). However, the focal spot remains near -10 μm, indicating the focal length does not change significantly with the axial separations. In the y-z plane, the focal spot remains near 10 μm and retains its shape, but decreases rapidly in intensity (e)-(h). The spot near -10 μm is the actual focal length, while the bright spot at the bottom half of the image is just the outline of the first metasurface. The simulated design is the same as in supplement S1, and has an in plane displacement of 4 $\mu$m. The axial separation is represented by the variable h, and the dashed white lines represent the locations of the two metasurfaces comprising the Alvarez lens.

In experiment, the metasurface near the objective remained stationary while the metasurface near the illumination source was translated backwards to increase the separation. The axial displacement slightly decreased the focal distance of the lens along both the x and y axes of the lens, but the shift is not appreciable, as seen in the theoretical analysis (Fig S7). However, the effect on the focal spot size was not deterministic, showing large spikes in one set of data and a gradual increase in the other.

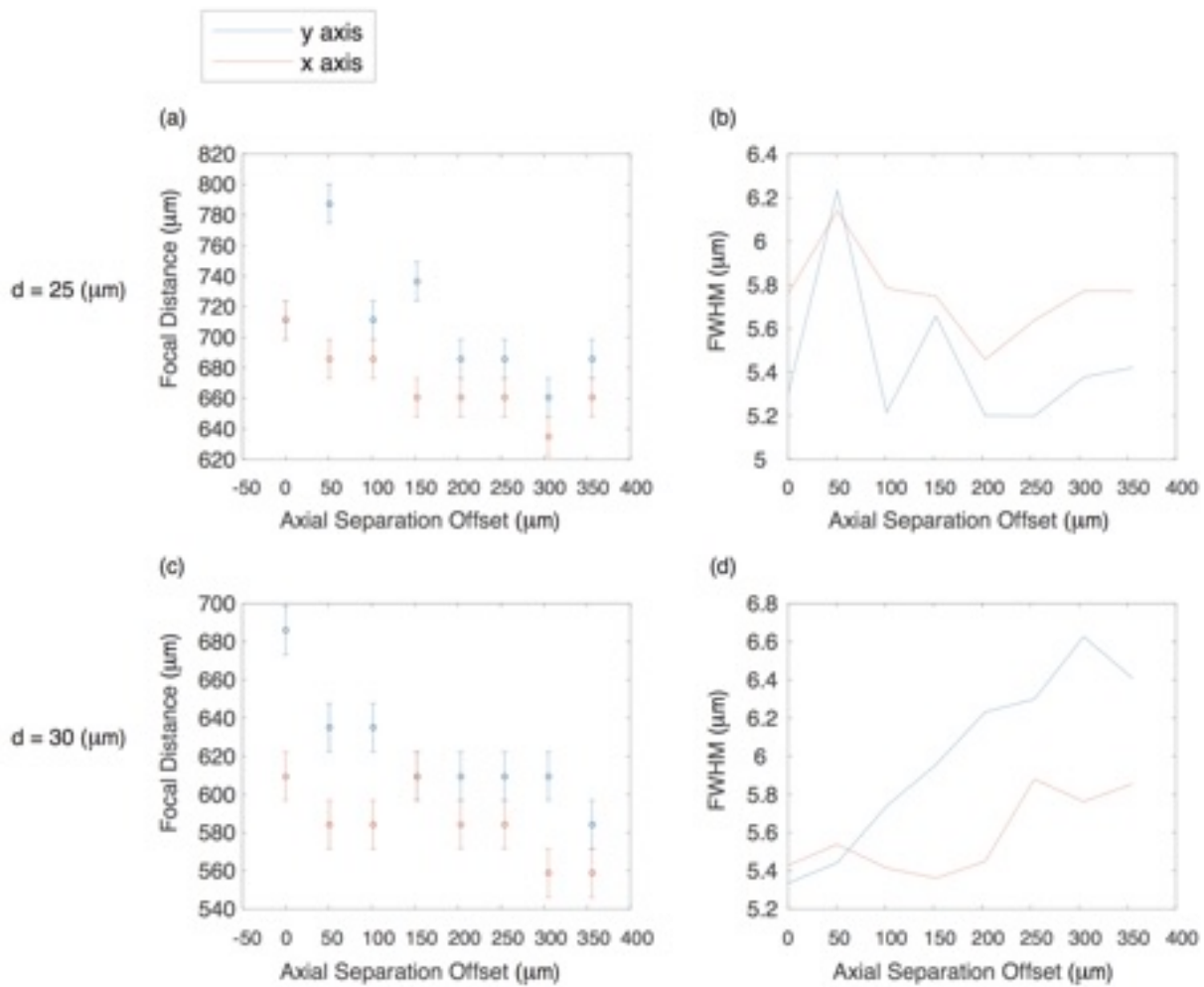

Fig. S7: Experimental Alvarez lens performance for separations along the optical axis. (a), (c) focal distances for an Alvarez lens with 25 and 30 μm of transverse displacement d respectively. As the axial separation increases, the lenses in both cases displayed a slight decrease in focal length. The axial displacement is not absolute, and can be thought of as an offset of some finite distance, as we cannot measure the actual distance with high accuracy. (b), (d) show the effect of the axial separation on the focal length of the Alvarez lens for 25 and 30 μm of transverse displacement respectively. Points in red and blue represent data taken from the y and x axes respectively. Error bars represent the mechanical error associated with our translation stage.

**S8: Cubic Image Retrieval**

In order for the cubic imaging system to provide useful images, the initial image must be post-processed by deconvolution of the cubic point spread function (PSF) from the initial image[1]. In order for the cubic phase plate to be useful in controlling chromatic aberrations, the PSF must be invariant over the wavelength range of interest. This is not possible in general for highly chromatic optical elements such as metasurfaces, but the metasurface cubic phase plate does satisfy this criterion for 633 nm and 532 nm illumination.

We quantify this invariance by calculating the modulation transfer function (MTF) of our experimentally measured PSFs using a two dimensional Fourier transform, shown in Fig S8 for the cubic elements and Fig S9 for the quadratic elements. The MTF gives the magnitude response of the system found by taking the magnitude of the optical transfer function or the Fourier transform of the PSF. The figures are 1D slices of a corresponding 2D MTF, which we are justified in taking a 1D slice of due to the rectangular separability of the phase function[2]. As shown in Figs S8 and S9, the cubic phase plate under green and red illumination exhibits very similar MTFs for a range of positions along the optical axis while the quadratic lens fails to do so. Notably for the cubic MTFs, the positions of the peaks and troughs are similar for low frequency components while this is not the case for the quadratic elements. Using the knowledge of our experimental PSF and MTF, a frequency domain filter can be constructed using a least squares optimization routine[3].

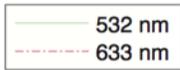

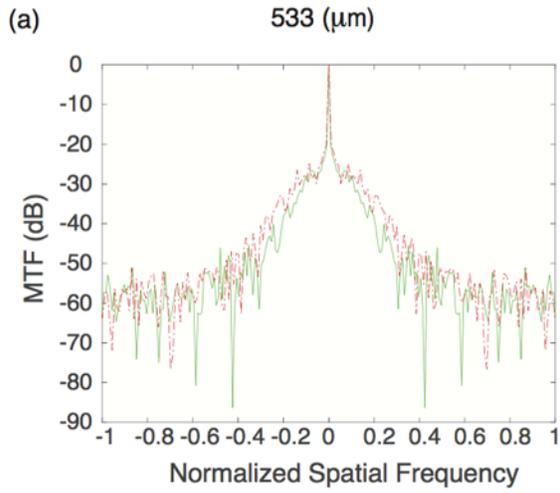
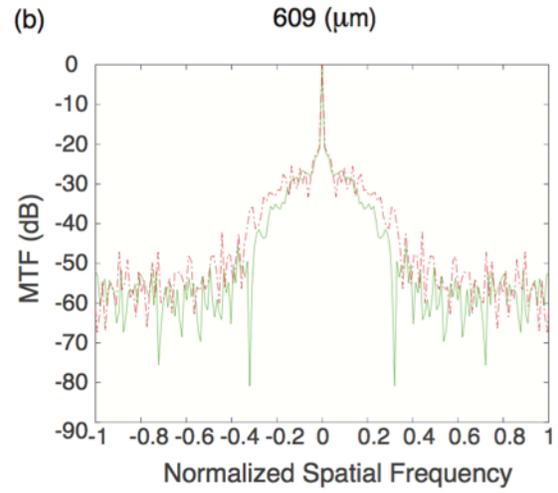
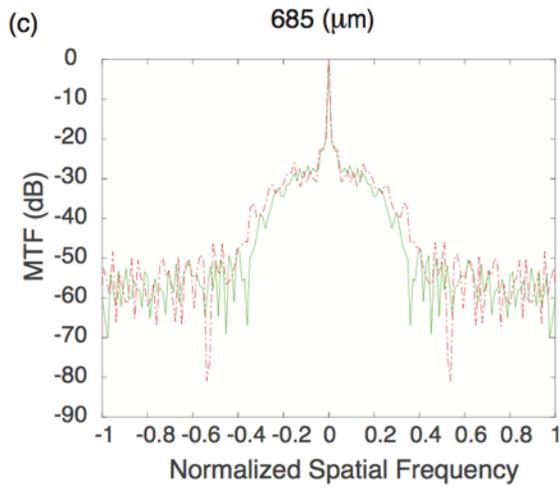
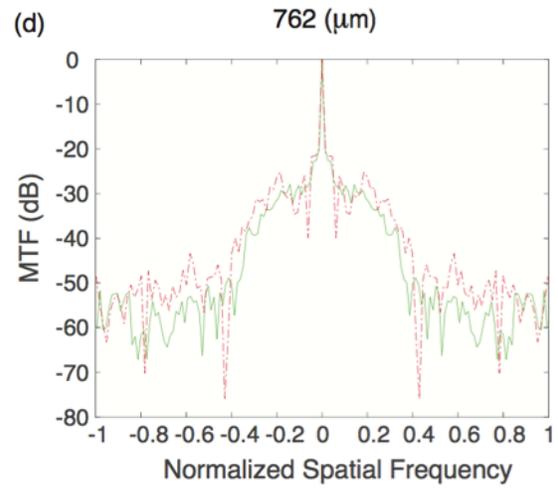
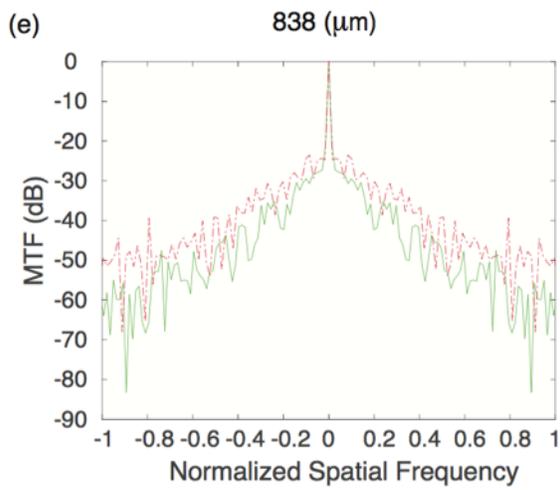
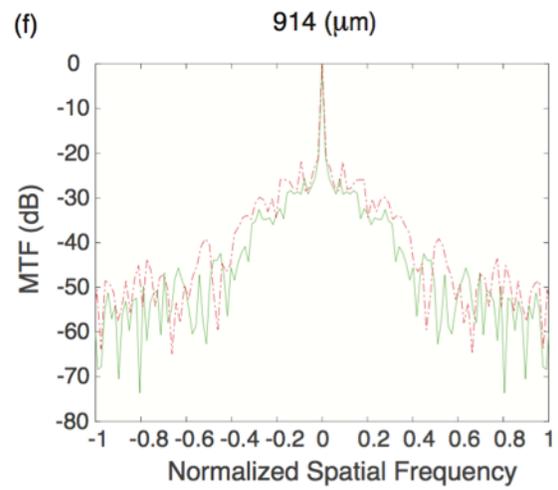

Fig. S8: Modulation transfer functions of the cubic element. (a)-(f) show 1D slices of the MTF of the cubic element for a range of over 300 μm plotted against normalized spatial frequency for both red and green illumination. The MTFs for green (532 nm) and red (633 nm) are shown in solid and dotted lines respectively.

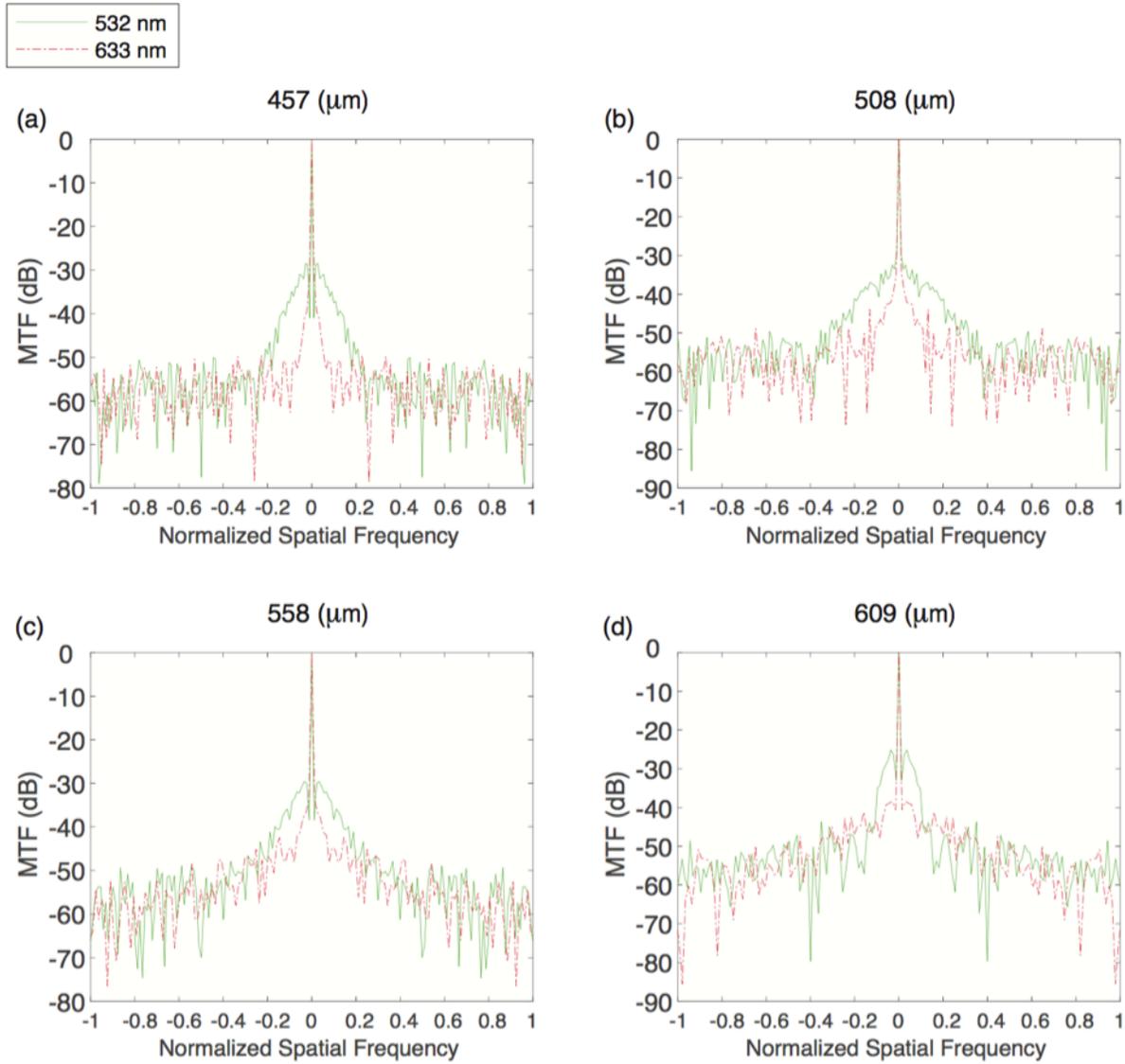

Fig. S9: Modulation transfer function of the 500 μm quadratic metasurface lens. (a)-(d) show 1D slices of the MTF of the quadratic element for a range of 150 μm plotted against normalized spatial frequency for both red and green illumination. The MTFs for green (532 nm) and red (633 nm) are shown in solid and dotted lines respectively.